# Multiple asymmetric couplings induced unconventional corner mode in topolectrical circuits

Hengxuan Jiang, [1] Xiumei Wang, [2*] Jie Chen, [1] and Xingping Zhou[3*]

[1] *College of Integrated Circuit Science and Engineering, Nanjing University of Posts and Telecommunications, Nanjing 210003, China*

[2] *College of Electronic and Optical Engineering, Nanjing University of Posts and Telecommunications, Nanjing 210003, China*

[3] *Institute of Quantum Information and Technology, Key Lab of Broadband Wireless Communication and Sensor Network Technology, Ministry of Education, Nanjing University of Posts and Telecommunications, Nanjing 210003, China*

*wxm@njupt.edu.cn*
*zxp@njupt.edu.cn*

We investigate the emergence of unconventional corner mode in a two-dimensional topolectrical circuits induced by asymmetric couplings. The non-Hermitian skin effect of two kinked one-dimensional lattices with multiple asymmetric couplings are explored. Then we extend to the two-dimensional model, derive conditions for the non-Hermitian hybrid skin effect and show how the corner modes are formed by non-reciprocal pumping based on one-dimensional topological modes. We provide explicit electrical circuit setups for realizing our observations via realistic LTspice simulation. Moreover, we show the time varying behaviors of voltage distributions to confirm our results. Our study may help to extend the knowledge on building the topological corner modes in the non-Hermitian presence.

## I. INTRODUCTION

Non-Hermiticity has brought about a plenty of interesting new phenomena in the field of the condensed matter, such as exceptional points (EPs) [1, 2], nodal rings [3-5], extensive localization of eigenstates [6-9], and unidirectional transport [10, 11]. One of the most iconic features of a non-Hermitian system is the non-Hermitian skin effect (NHSE) [7, 9, 12-30]. As we know, the presence of the non-Hermiticity can induce the breakdown of conventional bulk-boundary correspondence [31-34], which describe the connection between the band spectra under open boundary conditions (OBC) in the limit of infinite system size and those under periodic boundary conditions (PBC). The nonreciprocal Su-Schrieffer-Heeger (SSH) chain induced by giant atom also can show asymmetric zero states [35]. The non-Hermiticity may come from gain and loss effects [1, 2, 10], dissipations in open systems [3], or asymmetric couplings [12]. The non-Hermitian systems have been investigated in various synthetic platforms such as metamaterial [11], photonics [1-3, 26], mechanical [36], superconducting [37], and acoustic [22, 38, 39] systems. Recently, the electrical circuit become a suitable platform to show the many unconventional non-Hermitian topological phases due to the unprecedented flexibility in tuning the model parameters [6, 9, 25, 27, 28, 30]. Thus, the electrical circuit can help to expand NHSE to the higher dimensions gradually, which can provide fertile settings for the fascinating interplay between qualitatively distinct phenomena. In one-dimensional (1D) NHSE system, the wave functions are localized in the vicinity of the system boundaries under OBC, where the phase transition point for a finite system can be described by the non-Bloch topological invariant [31].

When the NHSE localization occurs in more than one direction, the higher-order NHSE system could be obtained, which arise more and more attention [15, 21, 40-43]. Much interesting physics, such as non-Hermitian Chern bands [33], defect induced the NHSE [44, 45], geometry dependent-skin effect [29], and hybrid skin-topological modes [9, 19], have been found in two and higher dimensional system. Although the general theory for the higher-dimensional skin effect has been established in Ref [14], the skin behavior in high-dimensional system can also be described via the topological invariant in the lower dimensional system [46].

In this work, we conduct the analysis of the unconventional corner mode in a two-dimensional (2D) topolectrical circuits induced by asymmetric couplings. Our model with odd-node configuration and central defects induces new topological characteristics and the distribution of modes becomes more intricate (e.g. the detachment of edge modes and skin modes). Firstly, we propose the 1D lattice with respect to two types of kinked structures. The NHSEs of 1D model are explored via non-Bloch winding number, spectra, and inverse participation ratios (IPR) [48]. Then, the kinked model is extended to a 2D lattice. Our coupled mode theory (CMT) calculations show the second-order skin effect (SOSE) in the 2D model. The skin corner weights are calculated to value the corner mode. It is revealed that the SOSE is originated from the skin pumping based on 1D topological modes. The conditions for appearance of skin effect behave the same as those in the 1D lattice. Furthermore, we present the scheme of a topolectrical circuit and perform LTspice simulation. The good agreements between the CMT calculation and LTspice simulation suggest the validity of the lattice model and analysis. Finally, we reveal the time behavior of the voltage profile based on CMT to confirm our results.

## Ⅱ. THE ONE-DIMENSIONAL NON-HERMITIAN SSH CHAIN CIRCUIT

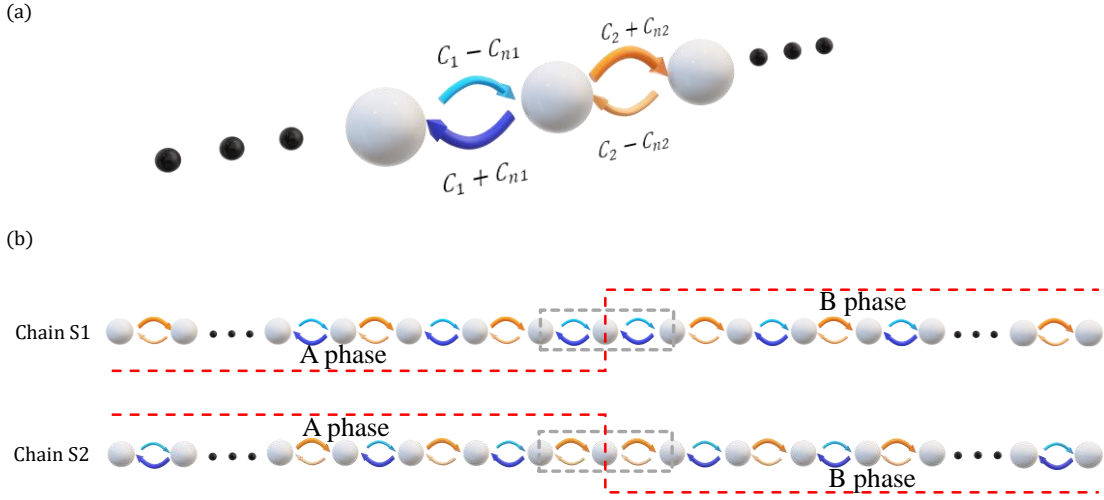

FIG. 1. A non-Hermitian SSH chain with asymmetric couplings. (a) The structure of 1D non-Hermitian SSH model. The intracell and intercell couplings are ($C_1 - C_{n1}$) and ($C_2 + C_{n2}$) in the forward direction and ($C_1 + C_{n1}$) and ($C_2 - C_{n2}$) in the backward direction, respectively. (b) The stucture of chain S1 and S2. As shown in the gray dashed boxes, two kinds of kink defects are induced.

We initially consider a generalized non-Hermitian SSH model depicted in Fig. 1(a). The non-Bloch Hamiltonian $\overline{H}(\beta)$ of the model is given by:

$$\overline{H}(\beta) = \begin{pmatrix} 0 & (C_1 - C_{n1}) + (C_2 - C_{n2})\beta \\ (C_1 + C_{n1}) + (C_2 + C_{n2})\beta^{-1} & 0 \end{pmatrix}. \tag{1}$$

Here, $\beta$ is the non-Bloch factor given by $\beta = re^{i\kappa} = e^{-\alpha}e^{i\kappa}$. The non-Bloch multiplication factor $\alpha$ denotes the exponential decay of the voltage. The non-Hermitian SSH model exhibits distinct intracell and intercell couplings in both forward and backward directions. The non-Hermiticity of the SSH chain circuit is induced by the $C_{n1}$ and $C_{n2}$. (i.e. for positive $C_1$, $C_{n1}$, $C_2$ and $C_{n2}$, $C_1 - C_{n1} < C_1 + C_{n1} < C_2 - C_{n2} < C_2 + C_{n2}$ in chain S1, $C_2 - C_{n2} < C_2 + C_{n2} < C_1 - C_{n1} < C_1 + C_{n1}$ in chain S2). In line with the SSH model notation, a perfect chain possesses two degenerated ground states – phases A and B [47], which correspond to the sublattice symmetry (i.e., chiral symmetry) with alternating double bonds and single bonds. In our non-Hermitian SSH model, two different asymmetric couplings are used in place of the single and double bonds shown in Fig. 1(b). In our work, we interconnect the chains with these two phases together, forming two distinct types of interfaces between phases A and B. This interconnection gives rise to two separate categories of topological excitations: kinked and anti-kinked (chain S1 and S2), as depicted by the grey dashed boxes in Fig. 1 (b).

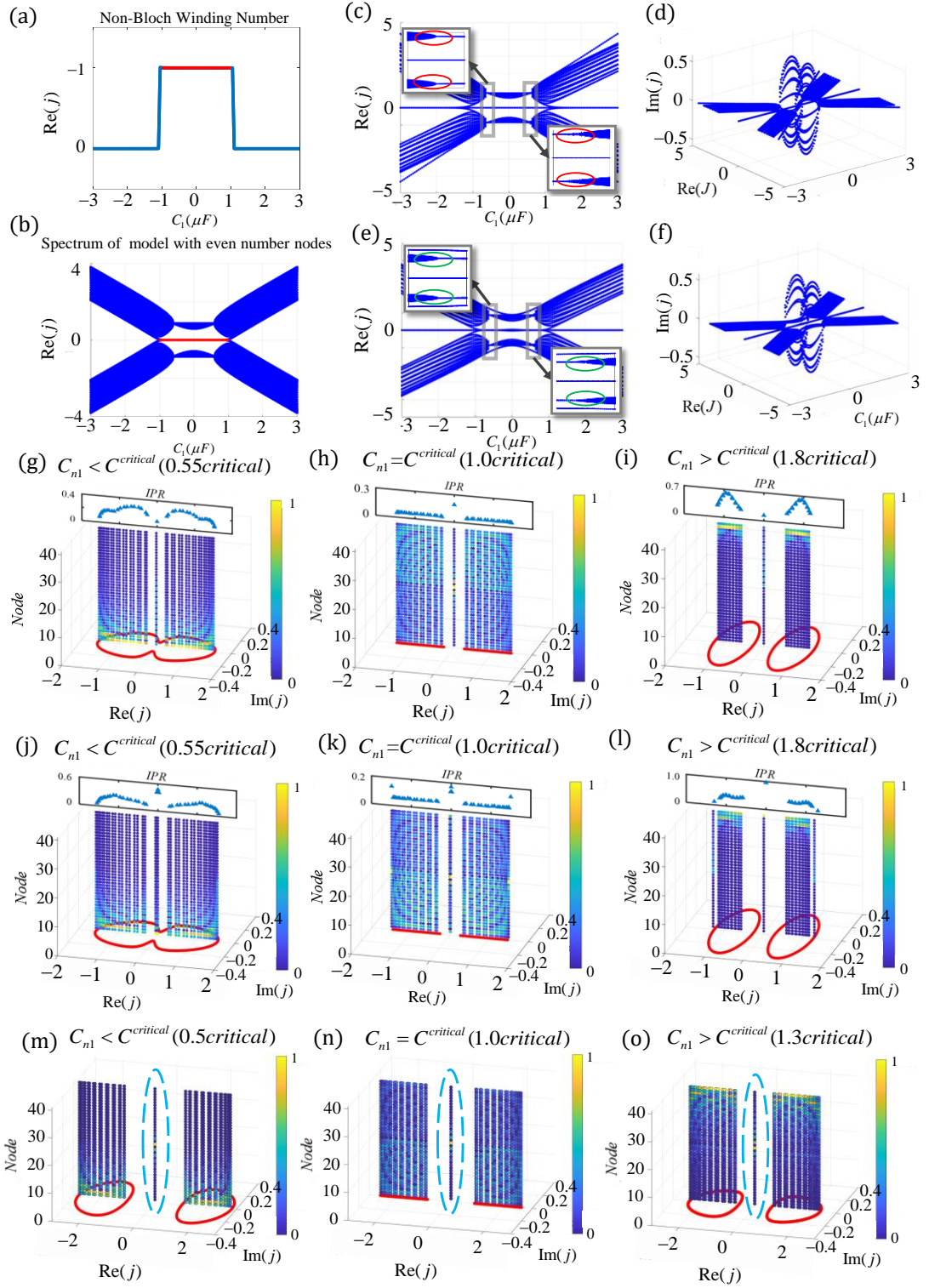

FIG. 2. (a) The non-Bloch winding number of non-Hermitian SSH model at $C_{n1} = 0.6\ \mu F$ and $C_{n2} = 0.5\ \mu F$. (b) The OBC admittance band dispersion of 40-node perfect non-Hermitian SSH model as a function of $C_1$ at $C_2 = 1\ \mu F$, $C_{n1} = 0.6\ \mu F$ and $C_{n2} = 0.5\ \mu F$. (c-d) The OBC admittance band dispersion of Chain S1 as a function of $C_1$ at $C_2 = 1\ \mu F$, $C_{n1} = 0.6\ \mu F$ and $C_{n2} = 0.5\ \mu F$. (e-f)

The OBC admittance band dispersion of Chain S2 as a function of $C_1$ at $C_2 = 1\,\mu F$, $C_{n1} = 0.6\,\mu F$ and $C_{n2} = 0.5\,\mu F$. The upper-left and lower-right images in Fig. 2(c, e) represent enlarged views of the admittance bands within the gray-bordered regions. (g-i) The OBC and PBC admittance spectra of Chain S1. (j-l) The OBC and PBC admittance spectra of Chain S2. Inset shows the IPR of eigenmodes. The parameters are set to $C_1 = 0.7\,\mu F$, $C_2 = 1\,\mu F$ and $C_{n2} = 0.5\,\mu F$. (m-o) The OBC and PBC admittance spectra of Chain S2. $C_1 = 2\,\mu F$, $C_2 = 1\,\mu F$ and $C_{n2} = 0.5\,\mu F$.

In the topolectrical circuit, the topological properties can be described by the Laplacian $J$. At a fixed frequency, the characteristic admittance $j_n$ of the Laplacian corresponds to the eigenenergies of the Hamiltonian in a real physical system. The voltage distribution corresponding to the $j_n$ simulates the eigenstates of the Hamiltonian.

Our analysis starts with the topological properties of the 1D non-Hermitian SSH models. Following the Kirchhoff's law, the circuit Laplacian can be written as

$$J(\beta,\omega) = (i\omega) \begin{pmatrix} \dfrac{1}{\omega^2 L_g} - C_2 & (C_1 - C_{n1}) + (C_2 - C_{n2})\beta \\ (C_1 + C_{n1}) + (C_2 + C_{n2})\beta^{-1} & \dfrac{1}{\omega^2 L_g} - C_2 \end{pmatrix}. \quad (2)$$

Here, $\beta$ is the non-Bloch factor given by $\beta = re^{i\kappa} = e^{-\alpha}e^{i\kappa}$. The non-Bloch multiplication factor $\alpha$ denotes the exponential decay of the voltage. Here, $i$ represents the imaginary number unit; $\omega$ represents the operating angular frequency of the circuit; and $L_g$ represents the value of the node parallel inductance. For simplicity, we represent $(i\omega)^{-1}J(\beta,\omega)$ as $H(\beta,\omega)$ and refer to it as the "normalized Laplacian" [28].

$$H(\beta,\omega) = \begin{pmatrix} \dfrac{1}{\omega^2 L_g} - C_2 & (C_1 - C_{n1}) + (C_2 - C_{n2})\beta \\ (C_1 + C_{n1}) + (C_2 + C_{n2})\beta^{-1} & \dfrac{1}{\omega^2 L_g} - C_2 \end{pmatrix}. \quad (3)$$

According to the principle of conventional bulk-boundary correspondence, conventional bulk topological index evaluated from the Bloch Hamiltonian can dictate topological modes in the finite system with OBCs. However, the conventional bulk-boundary correspondence become breakdown in non-Hermitian systems. The so-called non-Bloch winding number is introduced to fulfill the bulk-boundary correspondence, which is evaluated from the transformed Hamiltonian. It has been proven that non-Bloch winding number faithfully determines the number of topological modes in the non-Hermitian system [31]. Any parameters with nonzero non-Bloch winding number indicate the existence of zero-energy states in the non-Hermitian SSH model under OBC.

It is observed that when the model's node number is even, the non-Bloch winding number can reliably predict the appearance of zero-energy states (see the red line in Fig. 2(a, b)). However, when the model's node number is odd, the non-Bloch winding numbers no longer serve as predictors for the appearance of zero-energy states. As our model has an odd node number, the zero-energy states are no longer protected by non-Bloch winding numbers (shown in Fig. 2(c, e)).

Subsequently, we provide the admittance band structures for the defective Chain S1 and Chain S2. The OBC admittance band dispersion of Chain S1 as a function of $C_1$ at the resonant frequency is shown in Fig. 2(c-d). The OBC admittance band dispersion of Chain S2 as a function of $C_1$ at the resonant frequency is shown in Fig. 2(e-f). Due to the asymmetry in the coupling strengths of the model, the admittance bands also exhibit slight asymmetry, which is shown in the gray-bordered regions in Fig. 2(c, e). When the parameter $C_1$ varies in the range of [-3, 3], the coupling strengths $C_1 - C_{n1}$ scan across [-3.6, 2.4] and the coupling strengths $C_1 + C_{n1}$ scan across [-2.4, 3.6]. They are not symmetrically distributed in the parameter space, resulting in the left-right asymmetry.

We then analyze the evolution of NHSEs of Chain S1 and Chain S2. As is depicted in the Fig. 2(f-k), the red dots on the $\text{Re}(j) - \text{Im}(j)$ coordinate plane represent the admittance spectrum under PBC. Each column of blue to yellow dots represents the existence of eigenstates at complex admittance values given by the $\text{Re}(j)$ and $\text{Im}(j)$ coordinates of that column. For each eigenmode, the *z*-coordinate of the dots represents the spatial position along the chain, and the color of the dots at each location indicates the absolute value of the relative voltage amplitude at that node.

More specifically, the bulk voltage modes under OBC of the Chain S1 and S2 become localized at one of the boundary nodes (Fig. 2(f, h, i, k)), exhibiting with:

$$|V_x| \approx \beta^x = e^{-\kappa x} \tag{4}$$

where $\kappa$ represents [28]:

$$\kappa = -\frac{1}{2} \ln \left| \frac{(C_1 - C_{n1})(C_2 + C_{n2})}{(C_1 + C_{n1})(C_2 - C_{n2})} \right|. \tag{5}$$

When $\kappa$ reaches to zero, we denote the $C^{critcal}$:

$$C_{n1} = C^{critcal} = \frac{C_1 C_{n2}}{C_2}. \tag{6}$$

In this situation, the generalized Brillouin zone (GBZ) [31] transforms into the conventional Bloch zone of a unit circle in the complex plane, leading to the disappearance of the extensive localization in the voltage eigenstates shown in Fig. 2(g, h). Conversely, for any nonzero value of the inverse localization length $\kappa$, we can see the extensive localization of voltage eigenmodes. The $\kappa$ takes positive values when $C_{n1} > C^{critical}$, indicating that all voltage eigenmodes of both Chain S1 and S2 are localized at the left boundary (Fig. 2(f, i)). The $\kappa$ takes negative values when $C_{n1} < C^{critical}$. At this time, all voltage eigenmodes of both Chain S1 and S2 are localized at the right boundary (Fig. 2(h, k)).

Then, we conduct the analysis of the zero-energy states in Chain S1 and S2. We first look at the case when NHSE vanishes at $C_{n1} = C^{critical}$. For Chain-S1, depicted in Fig. 2(g), there is one zero-energy state identified and the field distribution indicates that it is induced by the domain wall of the kink. However, for Chain-S2, there are one zero-energy state, with its field distributions depicted in Fig. 2(j). It can be found that the zero-energy state shown in Fig. 2(j) is induced by the suspended nodes at the boundary of the non-Hermitian SSH chain. Then we turn to the case when NHSE becomes notable at $C_{n1} < C^{critical}$ and $C_{n1} > C^{critical}$. For both Chain S1 and S2, the zero-energy states are localized at the left boundary at $C_{n1} < C^{critical}$ shown in Fig. 2(f, i).

Conversely, the zero-energy states are localized at the right boundary at $C_{n1} > C^{critical}$ shown in Fig. 2(h, k).

In our model, the localization of the bulk and edge modes are respectively subjected to the NHSE and isolated site induced by one lattice with odd number, resulting in the detachment between the edge and skin modes. As depicted in the blue circle of Fig2. (m-o), it is noteworthy that the zero-energy states consistently exhibit localization in the middle of the model. Concurrently, the bulk modes are subject to the influence of NHSE.

To quantitatively characterize the localization property of our 1D model under OBC, we provide their IPRs given by:

$$IPR_n = \sum_{i=1}^{L} |V_i^n|^4 \tag{7}$$

where the summation is performed over all sites $(1, L)$ of the Chain and the eigenmode profiles $V^n$ are normalized by the condition $\sum_{i=1}^{L} |V_i^n|^2 = 1$. The value $IPR \to 0$ correspond to delocalized excitations, while the value $IPR \to 1$ indicate tight localization [48].

As depicted in the inset of Fig. 2(f-k), the IPRs of both Chain S1 and S2 get larger at $C_{n1} < C^{critical}$ and $C_{n1} > C^{critical}$ than that at $C_{n1} = C^{critical}$, which implies that the voltage distributions of eigenstates are gradually localized with nonzero $\kappa$. It should be noted that the field localization of zero-energy states in Chain S2 is more pronounced than that of Chain S1.

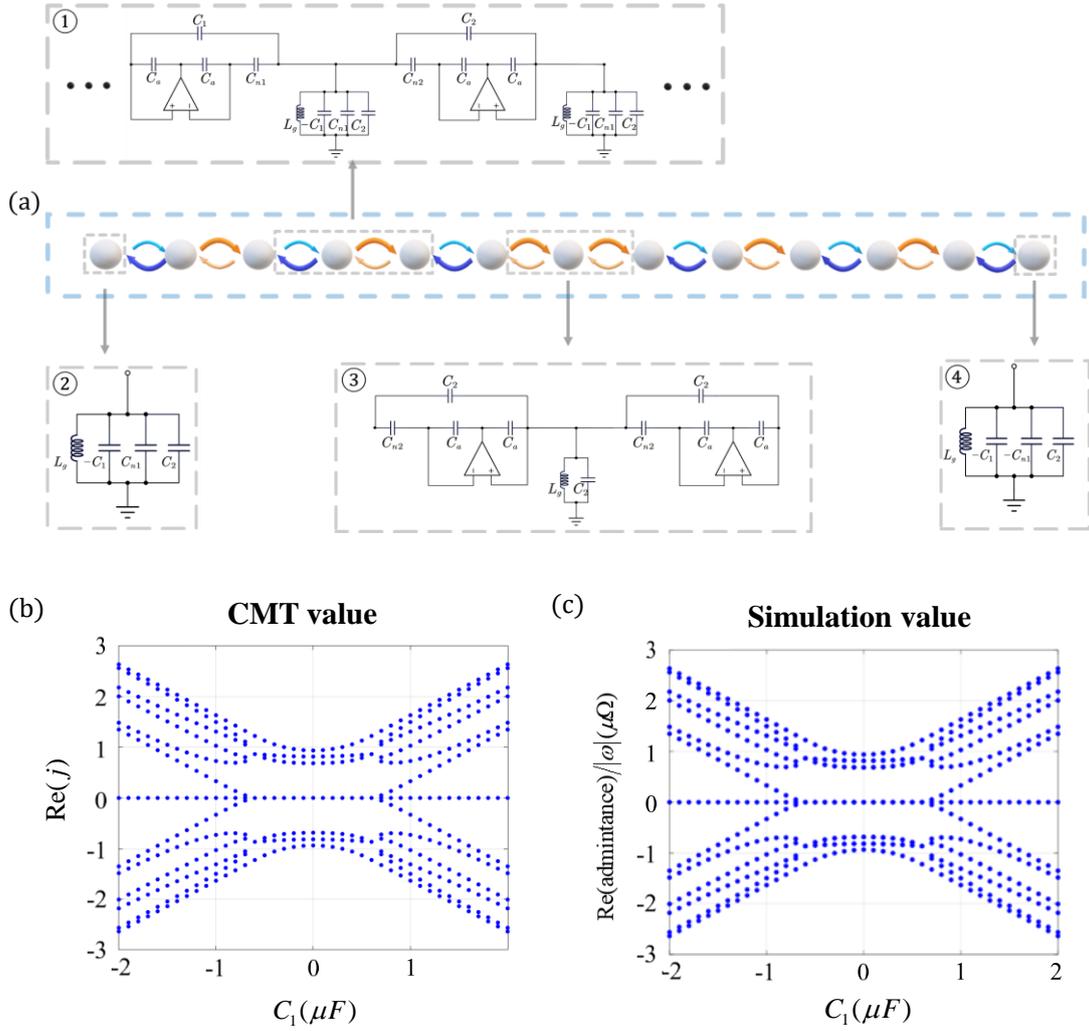

FIG. 3. The 1D topolectrical circuit with chain S2 type defect. (a) A 13-node non-Hermitian circuit chain with the defect of Chain S2 type. (b) The OBC admittance spectrum of the 13-node non-Hermitian SSH model with Chain S2 type defect by CMT. (c) The OBC admittance spectrum of 13-node circuit with Chain S2 type defect simulated by LTspice. The parameters are set to $C_1 \in [-2,2]$, $C_2 = 1\ \mu F$, $C_{n1} = 0.6\ \mu F$, $C_{n2} = 0.5\ \mu F$.

As shown in Fig. 3(a), we construct a 13-node 1D topolectrical circuit with Chain S2 defects. One unit cell of the circuit is marked with a dashed gray rectangle (i). The two terminal nodes are marked with the rectangle (ii) and (iv). Compared with the internal nodes, the terminal nodes (ii) and (iv) are suspended due to the lack of an asymmetric intercell coupling. The central defective cell is marked with the rectangle (iii). Compared with the unit cell (i), the asymmetric couplings in defective cell (iii) remain the same in both forward and backward direction. In the circuit chain, each node is grounded with capacitors and inductors. Specially, to remain the diagonal elements uniform throughout the Laplacian (Eq. 2), the grounding of nodes in (ii), (iii) and (iv) require distinct configuration from the nodes in (i). The asymmetric intercell and intracell couplings are realized through Negative Impedance Converters with current inversion (INICs).

To test the correctness of our circuit, we perform simulations via LTspice . In our simulation, we reconstruct the Laplacian matrix and the admittance spectra shown in Fig. 3(c). One way to obtain the Laplacian matrix is to inject single-frequency current sources into the circuit and measure the voltage responses. Then, we can get the Laplacian by:

$$J = IV^{-1}. \tag{8}$$

The principal diagonal element of the circuit Laplacian Eq. 2 is non-zero, and principal diagonal element of the Hamiltonian Eq. 1 is zero. To unify these two forms, the Laplacian principal diagonal elements need to be zero when the circuit is operating. Thus, we can get:

$$\frac{1}{\omega^2 L_g} - C_2 = 0. \tag{9}$$

According to Eq. 9, the frequency of the excitation current source is given by $f_r = \omega_r/2\pi = 1/2\pi\sqrt{L_g C_2}$. The details are provided in the Appendix A. Then, we calculate the theoretical admittance spectrum dispersion of a 13-node non-Hermitian SSH model with Chain S2 defects by CMT in Fig. 3(b). As we can see in Fig. 2(d) and Fig. 3(b), a reduction in the number of nodes has a limited effect on the dispersion of admittance spectrum for our model. The good agreement between the CMT results (Fig. 3(b)) and the LTspice results (Fig. 3(c)) demonstrates the validity of our model. We perform the simulation of zero-energy states in the 1D circuit via LTspice (see Appendix B), which exhibits consistency with our CMT calculation in Fig. 2(l).

## Ⅲ. THE TWO-DIMENSIONAL NON-HERMITIAN SSH NET CIRCUIT.

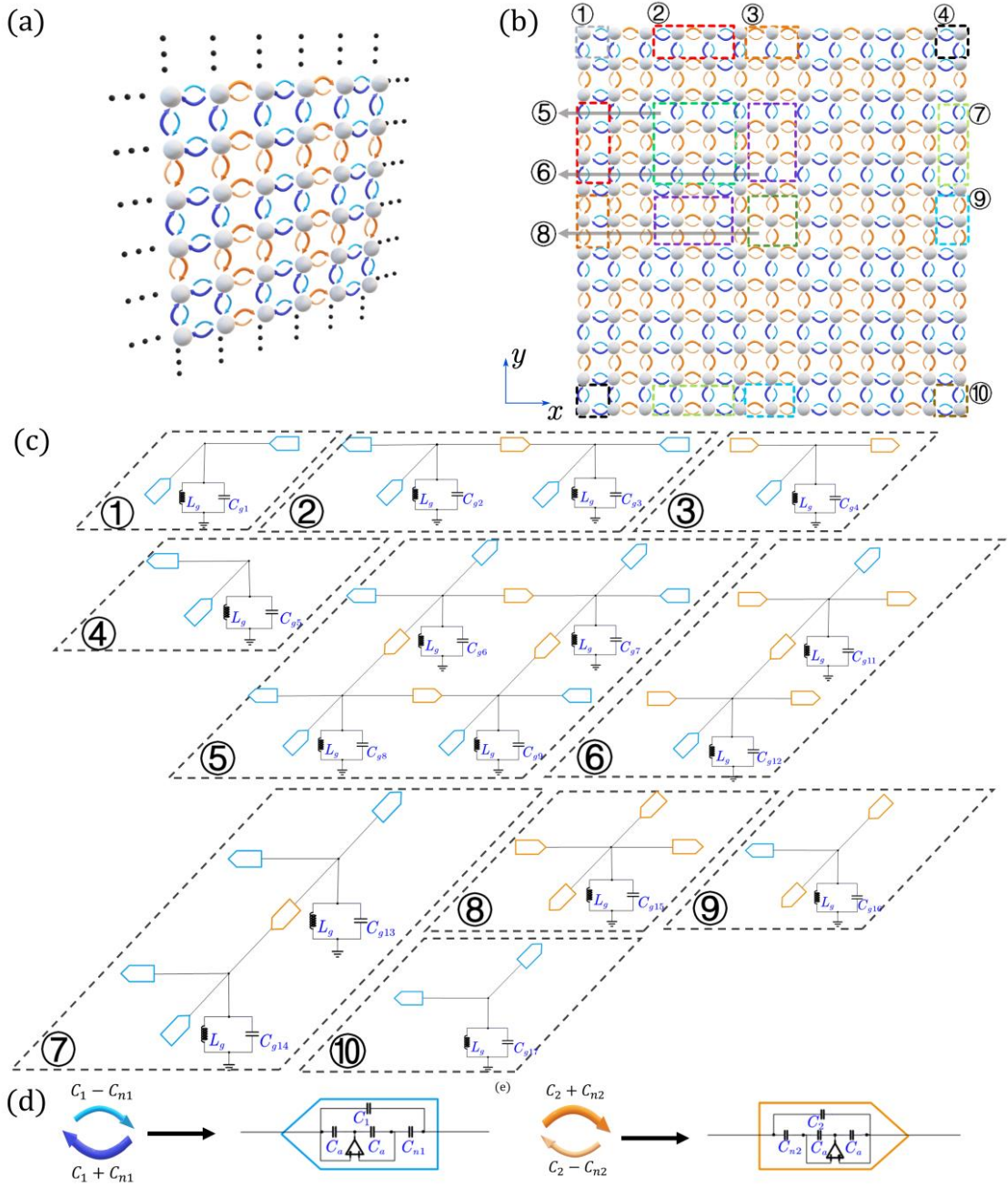

FIG. 4. The 2D non-Hermitian SSH model. (a) Layout of a 2D non-Hermitian SSH model under PBC. The intercell hopping couplings are ($C_1 - C_{n1}$) and ($C_2 + C_{n2}$) in the forward direction and down direction, while ($C_1 + C_{n1}$) and ($C_2 - C_{n2}$) in the backward and up direction, respectively. (b) Layout of a 2D non-Hermitian SSH model with defect under OBC. The structure of Chain S2 is induced in both $x$ and $y$ direction. (c) The structure of 2D non-Hermitian circuit. Each node and asymmetric hopping couplings are replaced with capacitors, inductors, and negative impedance amplifiers. (d-e) Two kinds of asymmetric coupling realized using INICs.

We extend our 1D non-Hermitian SSH model to 2D non-Hermitian SSH model under PBC, as shown in Fig. 4(a). The intracell and intercell hopping couplings are $C_1 - C_{n1}$ and $C_2 + C_{n2}$ in the forward direction and down direction, while $C_1 + C_{n1}$ and $C_2 - C_{n2}$ in the backward and up direction, respectively. We can write the normalized Laplacian matrix of the 2D circuit:

$$H(\beta_x, \beta_y, \omega) = \begin{pmatrix} \frac{1}{\omega^2 L_g} - C_2 & (C_1 - C_{n1}) + (C_2 - C_{n2})\beta_x & 0 & (C_1 - C_{n1}) + (C_2 - C_{n2})\beta_y \\ (C_1 + C_{n1}) + (C_2 + C_{n2})\beta_x^{-1} & \frac{1}{\omega^2 L_g} - C_2 & (C_1 - C_{n1}) + (C_2 - C_{n2})\beta_y & 0 \\ 0 & (C_1 + C_{n1}) + (C_2 + C_{n2})\beta_y^{-1} & \frac{1}{\omega^2 L_g} - C_2 & (C_1 + C_{n1}) + (C_2 + C_{n2})\beta_x^{-1} \\ (C_1 + C_{n1}) + (C_2 + C_{n2})\beta_y^{-1} & 0 & (C_1 - C_{n1}) + (C_2 - C_{n2})\beta_x & \frac{1}{\omega^2 L_g} - C_2 \end{pmatrix}.$$

(10)

Here, $\beta_{x,y}$ is the non-Bloch factor given by $\beta_{x,y} = re^{i\kappa_{x,y}} = e^{-\alpha} e^{i\kappa_{x,y}}$. The non-Bloch multiplication factor $\alpha$ denotes the exponential decay of the voltage. Then, we induce defects into our 2D non-Hermitian SSH model. The Chain S2 type defects are induced in both the *x* and *y* directions shown in Fig. 4(b). To realize the 2D non-Hermitian SSH model with electrical circuits, we replace each node and asymmetric coupling with electrical components shown in Fig. 4(c). We turn to analyze the configuration of electrical components. The grounded circuit Laplacian under OBC can be written as:

$$\begin{cases} J_{ab} = Y_{ab} & (a \neq b) \\ J_{ab} = Y_{a0} + \sum_{n=1}^{N} Y_{an} & (a = b) \end{cases}. \quad (11)$$

Here, *a* and *b* represent the index of the nodes. The $N$ represents the total number of nodes in the model. The $J_{ab}$ represents the elements at corresponding positions (*a*, *b*) in the circuit Laplacian. $Y_{ab}$ given by:

$$Y_{ab} = i\omega C_{ab} - \frac{i}{\omega L_{ab}} \quad (12)$$

represents the admittance between node *a* and *b*. The $L_{ab}$ represents the inductance value between nodes *a* and *b*. The $Y_{a0}$ denotes the admittance between node *a* and the ground [49].

Obviously, the non-diagonal terms of the Laplacians are determined by the asymmetric couplings between nodes (see Eq. 11 at $a \neq b$). Each diagonal term is related to the couplings with adjacent nodes and the grounding components at this node (see Eq. 11 at $a = b$). In our circuit design, a significance goal is to eliminate the diagonal terms in Eq. 10. At the fixed work frequency, we can eliminate the diagonal terms by configuring the grounded capacitances $C_g$ carefully with Eq. 11 and Eq. 12 (details in Appendix C).

Due to complexity induced by the 2D circuit, we divide the entire OBC circuit into 10 unit cells shown in the dashed rectangles ①-⑩ in the Fig. 4(b), whose specific circuit implements are shown in Fig.4 (c). The serial numbers ①-⑩ in the Fig. 4(b, c) show the correspondence between the 2D topolectrical circuit and the non-Hermitian SSH model. The corner cell ①, ④ and ⑩ contains two asymmetric couplings. The terminal cells ③ and ⑨ consist of three asymmetric couplings, including ② and ⑦ with five couplings. In the central defective cell ⑧, the four couplings of the node are the same. It can be observed that the nodes in the rectangles of same color exhibit structural symmetry, which gives them the same configurations as is depicted in Fig. 4(b). The specific expressions for the capacitance $C_{g1}$ - $C_{g16}$ are provided in the Appendix D. Two types of asymmetric couplings are implemented using INICs shown in Fig. 4(d-e).

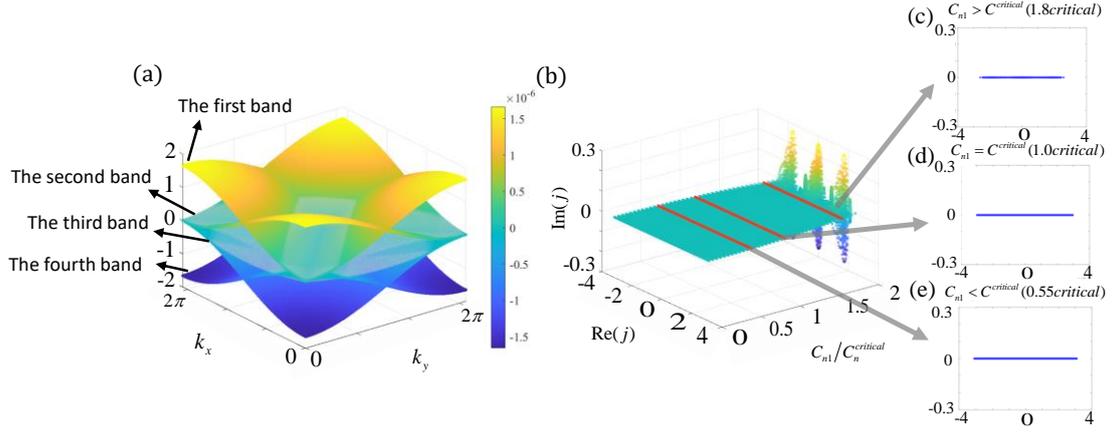

FIG. 5. The admittance band diagram of the 2D non-Hermitian SSH model. (a) 2D complex admittance-band structure numerically calculated from the Laplacian under PBC at $C_1 = 0.7\ \mu F$, $C_2 = 1\ \mu F$, $C_{n1} = 0.6\ \mu F$, $C_{n2} = 0.5\ \mu F$. (b) The OBC admittance spectra of 2D model as a function of $C_{n1}/C_n^{critical}$ with $C_1 = 0.7\ \mu F$, $C_2 = 1\ \mu F$, $C_{n2} = 0.5\ \mu F$. (c-e) The enlargement of the red portion shown in the figure.

We calculate the numerical results in the Brillouin Zone (BZ), producing the 2D complex admittance band structure depicted in the Fig. 5(a). In this diagram, the blue to yellow region represents the real admittance band structure of the 2D model. We refer to the four bands as the first band, the second band, the third band and the fourth band from top to bottom along the $z$-axis. The first and second bands are degenerate at $(0,\pi)$, $(\pi,0)$, $(\pi,2\pi)$ and $(2\pi,\pi)$. Similarly, the third and fourth bands are degenerate at $(0,\pi)$, $(\pi,0)$, $(\pi,2\pi)$ and $(2\pi,\pi)$. The second and third bands degenerate at the center and four corners in BZ.

Next, we proceed to analyze the admittance band structures under OBC. We first construct a 2D model, which consists of 1681 nodes arranged in a grid of 41 rows and 41 columns and forms Chain S2 type defects in both $x$ and $y$ directions. The 41×41 2D model only differs from the model shown in Fig. 4(b) in the number of nodes, and their defective cells are the same, which is already shown in the Fig. 4(c). Then, we calculate the dispersion of the admittance band as a function of $C_{n1}/C_n^{critical}$ for the 2D model at the resonant frequency, as depicted in Fig. 5(b). The imaginary part of the admittance spectrum doesn't appear until $C_{n1}/C_n^{critical}=1.85$ in Fig. 5(b). It is worth noting that the admittance band of the 2D model contains a zero-value imaginary part at $C_{n1}/C_n^{critical}=0.55$, $C_{n1}/C_n^{critical}=1.0$ and $C_{n1}/C_n^{critical}=1.8$ as is shown in the Fig. 5(c-e).

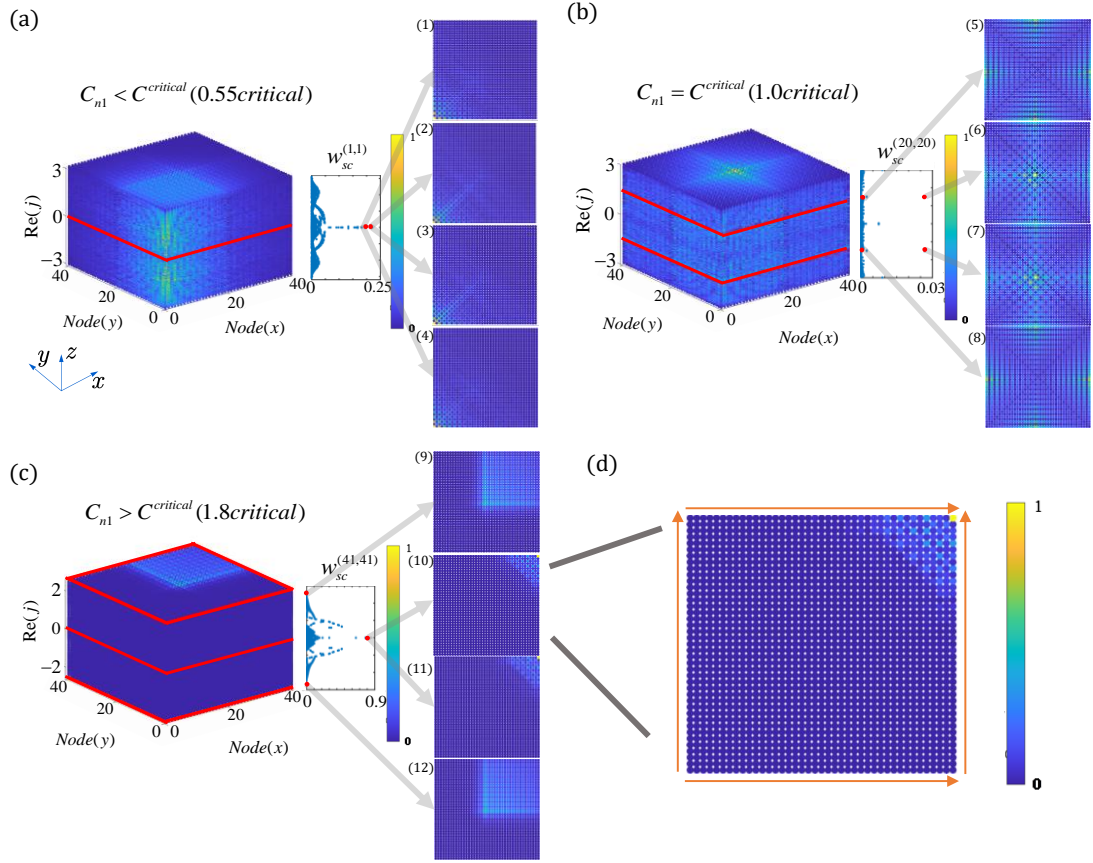

FIG. 6. The SOSE evolution of the 2D non-Hermitian SSH model under OBC. (a) The admittance spectra of 2D non-Hermitian SSH model under OBC at $C_{n1} < C^{critical}$ ($C_{n1} = 0.55C^{critical}$). (b) The admittance spectra of 2D non-Hermitian SSH model under OBC at $C_{n1} = C^{critical}$ ($C_{n1} = 1.0C^{critical}$). (c) The admittance spectra of 2D non-Hermitian SSH model under OBC at $C_{n1} > C^{critical}$ ($C_{n1} = 1.8C^{critical}$). The parameters are set to $C_1 = 0.7\ \mu F$, $C_2 = 1\ \mu F$ and $C_{n2} = 0.5\ \mu F$. (d) The direction of skin pumping.

We continue to consider the 41×41 2D model and turn to the evolution of corner mode in our 2D model. As shown in the Fig. 6(a-c), each plane parallel to the *x* and *y* axes represents the existence of eigenmodes under open OBC at the corresponding $\text{Re}(j)$ values (*z* coordinate). For each eigenmode, the coordinates (*x*, *y*) of the dots represent the spatial position on the 2D grid, and the color of the dots at each node indicates the absolute value of the relative voltage amplitude. The images on the right side of Fig. 6(a-c) provide a more detailed illustration of the distribution of voltage eigenmodes within the red region.

To quantitatively characterize the SOSE under OBC, we introduce the skin corner weight $w_{sc}$ defined as:

$$w_{sc}^{\mathbf{r_c}} = \sum_{n,\mathbf{r}} |V_n(\mathbf{r})|^4 \exp(-|\mathbf{r}-\mathbf{r_c}|/\xi) \tag{13}$$

where the sums run over all eigenstates indexed by $n$, lattice sites $\mathbf{r}$, and corner positions $\mathbf{r_c}$. The second factor $\exp(-|\mathbf{r}-\mathbf{r_c}|/\xi)$, which decays exponentially away from the corners with decay length $\xi$, selects only the modes localized at the corners. It means that the larger the $w_{sc}$ of a 2D eigenmode is, the more the eigenstate will locate to a point. If $\xi$ is much smaller than the linear system size, i.e., $\xi \ll L$. For SOSE, $w_{sc}$ is finite and scales linearly with $L$ in 2D model [50]. We select the corner position (1, 1), (20, 20) and (41, 41) as $\mathbf{r_c}$ in Eq. 13 at $C_{n1} < C^{critical}$, $C_{n1} = C^{critical}$ and $C_{n1} > C^{critical}$ respectively. The results of $w_{sc}^{(1,1)}$, $w_{sc}^{(20,20)}$ and $w_{sc}^{(41,41)}$ are depicted in the insets of Fig. 6(a-c). The insets share the $z$-axis with the admittance spectra. By finding the largest number of $w_{sc}$ values, we successfully find the tightest corner-like localization among all the eigenstates shown in (2), (3), (6), (7), (10) and (11). The red dots represent corresponding $w_{sc}$ of the eigenstates.

In our CMT calculation, the bulk modes under OBC of the 2D model still exhibit extreme localization like that of the 1D model (Fig. 6(a-c)). When $C_{n1} = C^{critical}$, the SOSE disappears, resulting in a widespread distribution of voltages throughout the 41×41 grid (Fig. 6(b)). When $C_{n1} < C^{critical}$, the bulk modes become localized in the lower-left corner of the 2D model (Fig. 6(a)). When $C_{n1} > C^{critical}$, the bulk modes are localized in the upper-right corner of the 2D model (Fig. 6(c)).

Then, we investigate the zero-energy states of the 2D model. We first turn to the two corner modes (10) and (11) at $C_{n1} > C^{critical}$ shown in Fig. 6(d). It should be noted that they are degenerate at zero admittance. The nonreciprocal skin pumping leads to accumulation of boundary skin modes along the $x$ and $y$ direction in these two modes

(corner modes) [9]. We choose "skin pumping" to describe it graphically: In the skin effect, the voltage distribution is increasingly located toward the model boundary. The orange arrows represent the direction of skin pumping in Fig. 6(d). The direction of non-reciprocal pumping is indicated by the 1D zero-energy state of Chain S2 at $C_{n1} > C^{critical}$ shown in Fig. 2(k). The degenerate eigenstates (9) and (12) are originated from skin pumping indicated by the 1D non-zero states in Fig. 2(k). In the same way, at $C_{n1} < C^{critical}$, the degenerate states (2) and (3) are originated from zero-energy state and the degenerate states (1) and (4) are originated from non-zero states in Fig. 2(i). At $C_{n1} = C^{critical}$, the degenerate states (6) and (7) are originated from zero-energy state and the degenerate states (5) and (8) are originated from non-zero states in Fig. 2(j). Like the case of 1D model, the localization of eigenstates disappears and the SOSE vanishes at $C_{n1} = C^{critical}$. In conclusion, it is evident that the SOSE is originated from the non-reciprocal skin effect on the 1D topological state along the *x* and *y* directions. In Fig. 6(a, c), we can see that our model is consistent with the above conclusions.

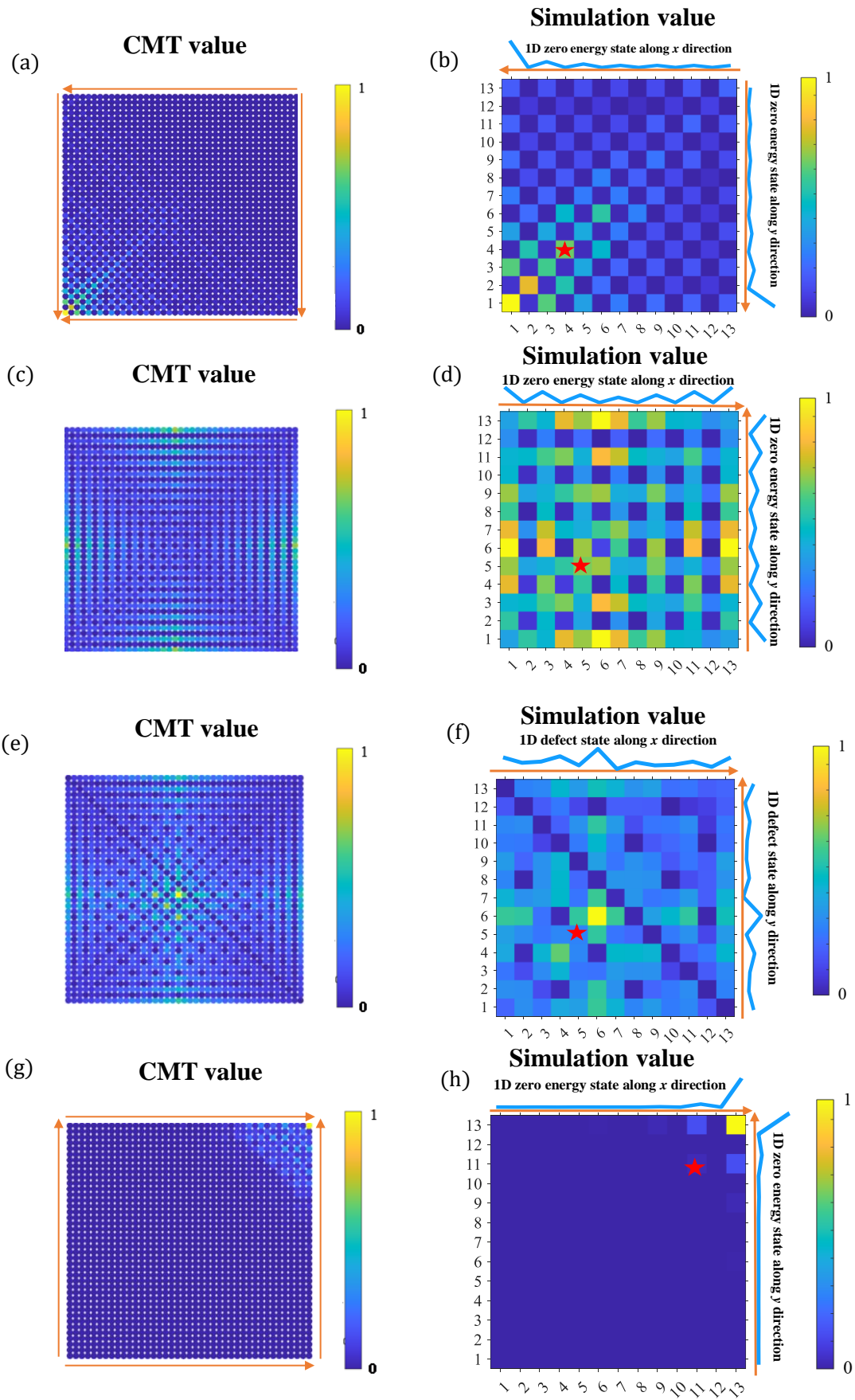

FIG. 7. CMT calculation and LTspice simulation of the 2D topological states. (a-b) The CMT value and the simulation value for the 2D corner states of the 2D non-Hermitian model at $C_{n1} < C^{critical}$ ($C_{n1} = 0.55 C^{critical}$). (c-f) The CMT value and the simulation results for the 2D topological states at $C_{n1} = C^{critical}$ ($C_{n1} = 1.0 C^{critical}$). (g-h) The CMT value and the simulation results for the 2D corner states at $C_{n1} > C^{critical}$ ($C_{n1} = 1.8 C^{critical}$) The parameters are set to $C_1 = 0.7\ \mu F$, $C_2 = 1\ \mu F$, $C_{n2} = 0.5\ \mu F$.

In our LTspice simulations, we choose the 13×13 structure, which is depicted in the Fig. 4(b). In this model, chain S2-type defects are induced in both the *x* and *y* directions. The specific circuit implementation is illustrated in Fig. 4(c-d). We select the eigenstates (2), (5), (6), and (10) in Fig. 6 for LTspice simulation. The details are provided in the Appendix A. As shown in Fig. 7(b, d, f, h), the blue lines represent LTspice simulation results of voltage distribution in the 1D topological circuit and the red asterisks represent the locations of the excitation. It is found that the voltage distributions along the *x* and *y* directions for the 2D eigenstates exhibit a clear similarity to the voltage distributions of the 1D eigenstates. It supports the opinion that the SOSE is originated from the non-reciprocal skin effect on the 1D topological state along the *x* and *y* directions. The orange arrows in Fig. 7(b, d, f, h) represent the direction of skin pumping. We can see that corner mode is originated from the skin pumping based on the zero-energy state in the 1D topolectrical circuit shown in Fig. 7(h).

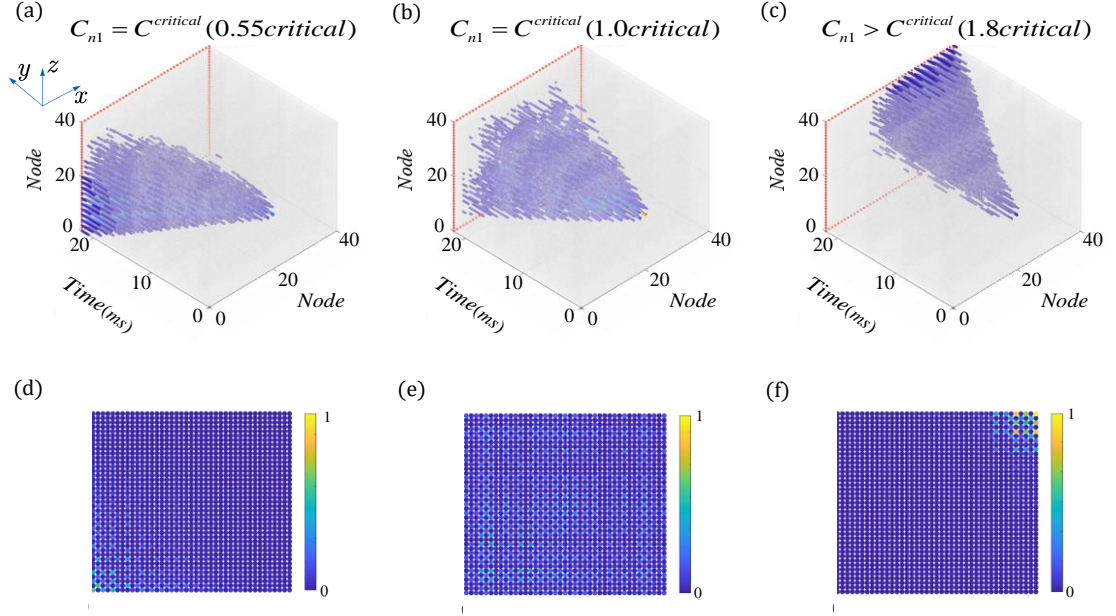

FIG. 8. The time varying behaviors of voltage distributions in the 41×41 2D non-Hermitian SSH model based on CMT. (a-c) Time-varying voltage under the injection of the central node. The *x-z* plane represents the spatial positions of different nodes in the 2D non-Hermitian model, while the *y*-axis corresponds to the temporal evolution at (d-f) Output profiles of voltage at different node in the 2D topolectrical circuit. The parameters are set to $C_1 = 0.7\ \mu F$, $C_2 = 1\ \mu F$, $C_{n2} = 0.5\ \mu F$.

Following the CMT, the voltage propagation in the 2D topolectrical circuit can be described by:

$$-j\frac{\partial}{\partial t}V_i = \beta_i V_i + \kappa_{i,i+1}V_{i+1} + \kappa_{i,i-1}V_{i-1} \tag{14}$$

where $V_i$ is the amplitude of the voltage in the *i*th node (here j =1, 2, . . . ,168, 169 according to our topolectrical circuit of N =169), and $\kappa$ is the coupling coefficient between the neighboring nodes. To reveal the time behaviors of voltage distributions in the 2D non-Hermitian model, we conduct calculations based on CMT. The distribution of time-varying voltage also exhibits strong localization phenomena. When $C_{n1} < C^{critical}$, the time-varying voltages localize in the lower-left corner of the 2D circuit (Fig. 8(a)). Conversely, when $C_{n1} > C^{critical}$, the time-varying voltages localize in the upper-right corner of the two-dimensional circuit (Fig. 8(c)). When $C_{n1} = C^{critical}$, the time-varying voltages maintain a widespread distribution Fig. 8(b).

## Ⅳ. CONCLUSION

To summarize, we successfully present a comprehensive description of the unconventional corner states in a 2D topolectrical circuits induced by asymmetric couplings. We first explore the NHSE of 1D lattice with respect to two types of kinked structures. We then design a 2D model with Chain S2-type defect. The SOSE is examined through our CMT calculation and the corner-like localization of the eigenstates is quantitatively characterized via skin corner weight. It is also derived that the appearance of the corner state is indicated by non-reciprocal pumping on 1D topological states. Furthermore, we implement the model to the realistic electrical circuits and perform the LTspice simulation of which the results successfully confirm with our CMT calculation. Moreover, the time behaviors of voltage profiles in the 2D model are revealed by CMT.

**Acknowledgements**
The authors thank for the support by NUPTSF (Grants No. NY220119, NY221055),

## APPENDIX A: SIMULATION PROPOSAL

*Reconstruct the Laplacian.* For the analysis of large-scale linear circuits in simulations, we employ the nodal analysis method. Through this approach, the admittance matrix of the circuit can be reconstructed by examining the voltage responses $V$ of each node to current injections $I$ under OBC.

$$I = JV. \tag{A1}$$

By injecting current excitations into the circuit and measuring the corresponding voltage responses, we can determine the circuit's admittance matrix using the following equation:

$$J_n^{m\times m} = \begin{pmatrix} I_1^1 & \cdots & I_1^{m\times m} \\ \vdots & \ddots & \vdots \\ I_n^1 & \cdots & I_n^{m\times m} \end{pmatrix} \times \begin{pmatrix} V_1^1 & \cdots & V_n^1 \\ \vdots & \ddots & \vdots \\ V_1^{m\times m} & \cdots & V_n^{m\times m} \end{pmatrix}^{-1} \tag{A2}$$

Here, $n$ represents the number of simulation iterations and $m \times m$ represents the lattice site of 2D grid. In the $i$-$th$ simulation, we can get vector $\begin{pmatrix} I_i^1 & \cdots & I_i^{m\times m} \end{pmatrix}$ and $\begin{pmatrix} V_i^1 & \cdots & V_i^{m\times m} \end{pmatrix}^T$. In this way, the number of simulations runs from 1 to n by index $i$. We can get matrix $I$ and $V$. Since only one node is injected with current at a time, matrix $I$ is an identity matrix. To obtain the reconstructed admittance matrix, we pre-multiply $V^{-1}$ by $I$.

*The single frequency excitation:* In accordance with the nodal analysis method in linear circuit analysis, all voltage and current quantities in the equations correspond to the magnitudes of sinusoidal AC signals at the single frequency. The frequency of source should be set to $f_r = \omega_r/2\pi = 1/2\pi\sqrt{L_g C_2}$. The diagonal terms of Laplacian vanish at $f_r$ to avoid the eigenvalues shift. During simulations, the excitations injected into the circuit are single-frequency AC signals. The voltage measurements in matrix V must retain the relative phase information with respect to the excitation current sources. Assuming an initial phase of zero for the injected current sources, the complex representation of matrix $V_i^j$ in terms of amplitude and phase ($v_m$ and $\varphi$) should be as follows:

$$V = \text{Im}(V_m e^{j\phi}) = iV_m \sin\varphi \tag{A3}$$

$$J = \frac{\text{Im}(V_m e^{j\phi})}{I_m} = iY_m \sin\varphi \tag{A4}$$

$$H = \frac{J}{j\omega} = \frac{Y_m \sin\varphi}{\omega}. \tag{A5}$$

Here, $V_m$ represents the amplitude of the single-frequency AC voltage, and $\varphi$ is the initial phase of the voltage signal. Given that our analysis focuses on the impedances of first-order components like capacitors and inductors, we only consider the imaginary part of their admittances. This is equivalent to taking the imaginary part of the admittance (i.e., susceptance), which effectively filters out the non-ideal effects of other circuit elements. The phase information of the response voltage signal can be easily obtained through AC analysis in software or by using a lock-in amplifier for measurement.

*Implementation of Negative Capacitance:* Negative capacitance can also be equivalently realized using INICs. However, the duality of capacitance and inductance allows us to employ an inductive equivalent for negative capacitance at the fixed frequency. In our simulations, we only care about single-frequency AC signals. Assuming the equivalent inductance value is L, we can simply calculate it by equating the impedance as follows:

$$X_c = \frac{1}{i\omega(-|C|)} = \frac{i}{\omega|C|}, \tag{A6}$$

$$X_{Leq} = i\omega L_{eq}. \tag{A7}$$

By setting $X_c$ equal to $X_{Leq}$, we can straightforwardly derive the equivalent inductance value:

$$L_{eq} = \frac{1}{\omega^2 |C|}. \tag{A8}$$

# APPENDIX B: THE SUPPLEMENTARY RESULTS OF SIMULATION

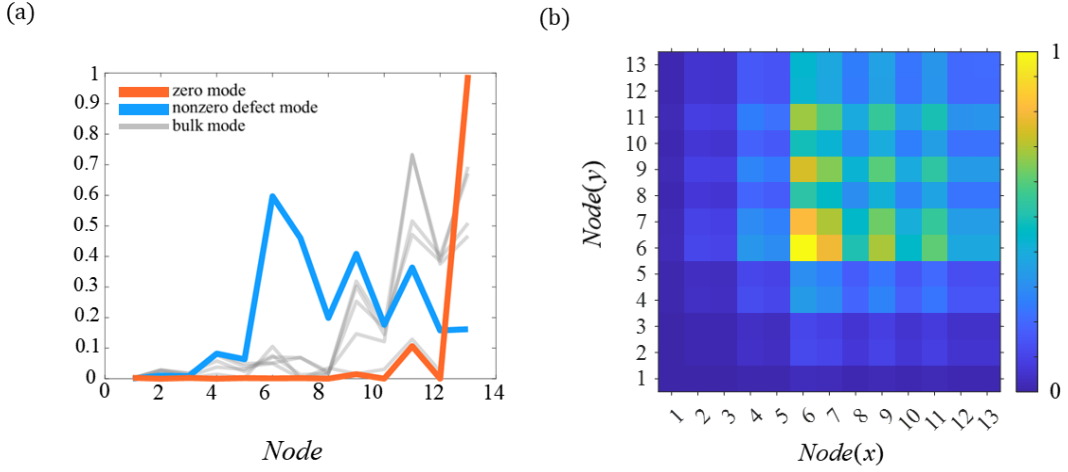

FIG. S1 The supplementary results of LTspice simulations for our 1D and 2D topolectrical circuit. (a) The voltage profiles of eigenmodes in 1D circuit. (b) The voltage profiles of nonzero defect mode in 2D circuit.

In our simulation of 1D topolectrical circuit, we get the zero mode and nonzero defect mode at $C_{n1} > C^{critical}$ shown in the orange and blue line in the Fig. S1(a), respectively. We can see that the zero mode exhibits the largest IPR among all the eigenstates in Fig. 2(k). This indicates the most pronounced localized voltage profile in our simulation via LTspice. For the simulation 2D circuit, we also find the 2D defect states shown in Fig. S1(b) corresponding to the degenerated states (9) and (12) in Fig. 6(c). It should be noted that the 2D defect state in Fig. S1(b) is originated from non-reciprocal pumping based on the 1D non-zero defect mode shown in the blue line of Fig. S1(a). This suggests the validity of our circuit strongly.

# APPENDIX C: THE METHOD OF ELIMINATE DIAGONAL TERM OF THE LAPLACIAN

To clearly explain the process in circuit design, we take a 5-node model with a central defect as an example.

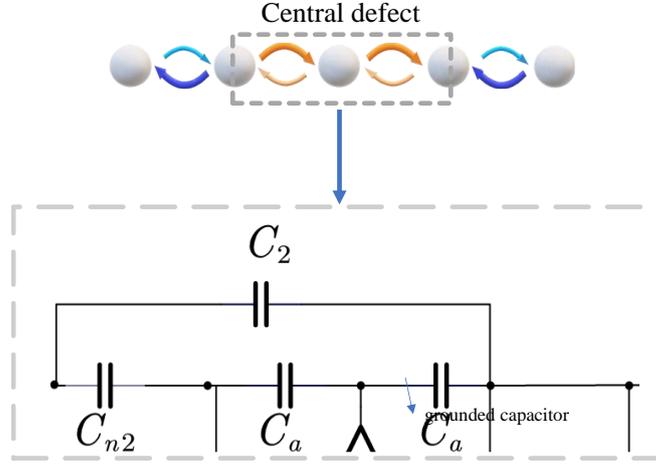

Fig. S2. A 5-node model with a central defect.

To simplify the notation, let $A = C_1 - C_{n1}$, $B = C_1 + C_{n1}$, $C = C_2 + C_{n2}$, $D = C_2 - C_{n2}$.

As is shown in the red terms of Eq. (C1), the off-diagonal circuit Laplacian matrix elements necessarily also appear with opposite sign as diagonal elements for the respective site [49].

$$\begin{bmatrix} \frac{1}{\omega^2 L_g} - C_2 - A & A & & & \\ B & \frac{1}{\omega^2 L_g} - C_2 - B - C & C & & \\ & D & \frac{1}{\omega^2 L_g} - C_2 - C - D & C & \\ & & D & \frac{1}{\omega^2 L_g} - C_2 - A - D & A \\ & & & B & \frac{1}{\omega^2 L_g} - C_2 - B \end{bmatrix} \quad (C1)$$

To eliminate the red term, we induce $\Delta_i$ by adding grounded capacitors at the circuit nodes (see Eq. (C2)) and set specific $\Delta_i$ to eliminate the red item (see Eq. (C3)).

$$\begin{bmatrix} \frac{1}{\omega^2 L_g} - C_2 - A + \Delta_1 & A & & & \\ B & \frac{1}{\omega^2 L_g} - C_2 - B - C + \Delta_2 & C & & \\ & D & \frac{1}{\omega^2 L_g} - C_2 - C - D + \Delta_3 & C & \\ & & D & \frac{1}{\omega^2 L_g} - C_2 - A - D + \Delta_4 & A \\ & & & B & \frac{1}{\omega^2 L_g} - C_2 - B + \Delta_5 \end{bmatrix} \quad (C2)$$

$$\begin{cases}\Delta_1=A\\ \Delta_2=B+C\\ \Delta_3=C+D\\ \Delta_4=D+A\\ \Delta_5=B\end{cases} \longrightarrow \begin{bmatrix} \frac{1}{\omega^2 L_g}-C_2 & A & & & & \\ B & \frac{1}{\omega^2 L_g}-C_2 & C & & & \\ & D & \frac{1}{\omega^2 L_g}-C_2 & C & & \\ & & D & \frac{1}{\omega^2 L_g}-C_2 & A & \\ & & & & B & \frac{1}{\omega^2 L_g}-C_2 \end{bmatrix} \quad (C3)$$

Let topolectrical circuit operate at a fixed angular frequency $\omega_r$,

$$\xrightarrow{\omega_r=1/\sqrt{L_g C_2}} \begin{bmatrix} 0 & A & & & & \\ B & 0 & C & & & \\ & D & 0 & C & & \\ & & D & 0 & A & \\ & & & & B & 0 \end{bmatrix} \text{(Consistent with the form of the Hamiltonian)} \quad (C4)$$

In a brief, eliminating the diagonal element of Laplacian is the key to our circuit design. Taking a principal diagonal element in Eq. (C1) as an example, we briefly summarize the ways of eliminating principal diagonal elements (shown in Fig. S3).

The principal diagonal element of the circuit Laplacian

| | $\frac{1}{\omega^2 L_g}-C_2$ | $-C-D$ |
|---|---|---|
| Why is this item induced? | By the resonators at circuit nodes. | By the asymmetric couplings with neighbor nodes. |
| How to eliminate? | Let topolectrical circuit operate at a fixed angular frequency. | Induce grounded capacitors at circuit nodes. |

Fig. S3. The ways of eliminating principal diagonal elements of the Laplacian. $C=C_2+C_{n2}$, $D=C_2-C_{n2}$.

# APPENDIX D: CAPACITOR VALUES OF THE 2D TOPOLETRICAL CIRCUIT

The specific expressions for the capacitance $C_{g1}$ - $C_{g16}$ in Fig. 4(c) are provided in Eq. (D1). The parameters are set to $C_1 = 0.7\ \mu F$, $C_2 = 1\ \mu F$, $C_{n1} = 0.63\ \mu F$, $C_{n2} = 0.5\ \mu F$ in our CMT calculations and simulations.

$$\begin{cases} C_{g1} = -2C_1 + 2C_{n1} + C_2 \\ C_{g2} = -2C_1 - C_{n2} \\ C_{g3} = -2C_1 + 2C_{n1} + C_{n2} \\ C_{g4} = -C_1 + C_{n1} - C_2 \\ C_{g5} = -2C_1 + C_2 \\ C_{g6} = -2C_1 - 2C_{n1} - C_2 - 2C_{n2} \\ C_{g7} = -2C_1 - C_2 \\ C_{g8} = -2C_1 + C_2 \\ C_{g9} = -2C_1 + 2C_{n1} - C_2 + 2C_{n2} \\ C_{g10} = -2C_1 - C_{n1} - 2C_2 - C_{n2} \\ C_{g11} = -C_1 + C_{n1} - 2C_2 + C_{n2} \\ C_{g12} = -2C_1 - 2C_{n1} - C_{n2} \\ C_{g13} = -2C_1 + C_{n2} \\ C_{g14} = -2C_1 \\ C_{g15} = -C_1 - C_{n1} - C_2 \\ C_{g16} = -2C_1 - 2C_{n1} + C_2 \end{cases} \quad (D1)$$